\documentclass[a4paper,preprint,superscriptaddress,preprintnumbers,nofootinbib]{revtex4-1}
\usepackage{graphicx}
\usepackage{subcaption}
\usepackage{float}
\usepackage{wrapfig}
\usepackage{bm}
\usepackage{color}
\usepackage{comment}
\usepackage{xcolor}
\usepackage[normalem]{ulem}
\usepackage{hyperref}
\usepackage{enumerate}
\usepackage{slashed}
\usepackage{mathtools}
\usepackage{cleveref}
\usepackage{ulem}
\usepackage{appendix}

\catcode`\@=11
\def\lsim{\mathrel{\mathpalette\@versim<}}
\def\gsim{\mathrel{\mathpalette\@versim>}}
\def\@versim#1#2{\vcenter{\offinterlineskip
\ialign{$\m@th#1\hfil##\hfil$\crcr#2\crcr\sim\crcr } }}
\catcode`\@=12

\usepackage{amsmath,amssymb}
\DeclareFontFamily{U}{mathx}{}
\DeclareFontShape{U}{mathx}{m}{n}{<-> mathx10}{}
\DeclareSymbolFont{mathx}{U}{mathx}{m}{n}
\DeclareMathAccent{\widehat}{0}{mathx}{"70}
\DeclareMathAccent{\widecheck}{0}{mathx}{"71}

\newcommand{\Slash}[1]{{\ooalign{\hfil/\hfil\crcr$#1$}}}

\DeclareMathOperator{\Tr}{Tr}

\DeclareMathOperator{\diag}{diag}

\usepackage{pict2e}
\makeatletter
\DeclareRobustCommand{\intprod}{%
  \mathbin{\mathpalette\int@prod{(0.1,0)(0.85,0)(0.85,0.7)}}%
}
\DeclareRobustCommand{\intprodr}{%
  \mathbin{\mathpalette\int@prod{(0.1,0.7)(0.1,0)(0.85,0)}}}

\newcommand{\int@prod}[2]{%
  \begingroup
  \sbox\z@{$\m@th#1+$}%
  \setlength\unitlength{\wd\z@}%
  \begin{picture}(1,1)
  \roundcap
  \polyline#2
  \end{picture}%
  \endgroup
}
\makeatother

\newcommand{\al}[1]{\begin{align}#1\end{align}}

\newcommand{\ov}{\over}
\newcommand{\nn}{\nonumber\\}
\newcommand{\tx}{\text}

\newcommand{\os}[2]{\overset{#1}{#2}{}}

\newcommand{\pn}[1]{\left(#1\right)}
\newcommand{\sqbr}[1]{\left[#1\right]}
\newcommand{\ab}[1]{\left|#1\right|}

\newcommand{\fn}[1]{\!\left(#1\right)}

\newcommand{\pa}[1]{\left(#1\right)\!{}}

\newcommand{\bs}{\boldsymbol}

\newcommand{\df}{\text{d}}

\newcommand{\mf}{\mathfrak}

\newcommand{\p}{\partial}

\newcommand{\pr}{\prime}

\newcommand{\ba}{\textbf{a}}
\newcommand{\bb}{\textbf{b}}
\newcommand{\bc}{\textbf{c}}
\newcommand{\bd}{\textbf{d}}

\definecolor{darkgreen}{rgb}{0,0.75,0}
\definecolor{darkred}{rgb}{0.75,0,0}
\definecolor{darkyellow}{rgb}{0.75,0.75,0}
\definecolor{darkcyan}{rgb}{0,0.75,0.75}
\definecolor{darkmagenta}{rgb}{0.75,0,0.75}

\newcommand{\sr}[2]{\stackrel{#1}{#2}}

\graphicspath{{./figs/}}
\newbox{\ORCIDicon}\sbox{\ORCIDicon}{\large
\includegraphics[width=0.8em]{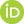}}

\usepackage{mathrsfs}
\usepackage[OMLmathsfit]{isomath}

\usepackage{bm}

\usepackage{braket}

\begin{document}

\title{
Spacetime and Planck mass generation from scale-invariant degenerate gravity
}
\author{Yadikaer \surname{Maitiniyazi}
\href{https://orcid.org/0009-0004-0826-1130}{\usebox{\ORCIDicon}}
}
\email{ydqem22@mails.jlu.edu.cn}
\affiliation{Center for Theoretical Physics and College of Physics, Jilin University, Changchun 130012, China}
\author{Shinya \surname{Matsuzaki}
\href{https://orcid.org/0000-0003-4531-0363}{\usebox{\ORCIDicon}}
}
\email{synya@jlu.edu.cn}
\affiliation{Center for Theoretical Physics and College of Physics, Jilin University, Changchun 130012, China}

\author{Kin-ya \surname{Oda}
\href{https://orcid.org/0000-0003-3021-1661}{\usebox{\ORCIDicon}}
}
\email{odakin@lab.twcu.ac.jp}
\affiliation{Department of Mathematics, Tokyo Woman’s Christian University, Tokyo 167-8585, Japan}

\author{Masatoshi \surname{Yamada}
\href{https://orcid.org/0000-0002-1013-8631}{\usebox{\ORCIDicon}}
}
\email{yamada@jlu.edu.cn}
\affiliation{Center for Theoretical Physics and College of Physics, Jilin University, Changchun 130012, China}

\begin{abstract}
We investigate a gravitational model based on local Lorentz invariance and general coordinate invariance. The model incorporates classical scale invariance, which forbids dimensionful parameters, and the irreversible vierbein postulate, which enables continuous degenerate limits of the vierbein, both at a specific scale. Through the dynamics of the system, we demonstrate the simultaneous emergence of the Planck mass and a curved spacetime background.
\end{abstract}

\maketitle
\section{Introduction}

The origin of spacetime and mass scales remains one of the deepest mysteries in theoretical physics. General relativity describes the macroscopic structure of spacetime but does not explain its fundamental origin or the emergence of the Planck scale. In quantum field theory, while mechanisms such as spontaneous symmetry breaking can organize physical scales, these scales are typically introduced externally, as in the case of the Standard Model. This limitation raises the question of whether a deeper framework could provide a unified explanation for the origin of both spacetime and mass scales, motivating the pursuit of quantum gravity as a potential path forward.

It is well known, however, that quantizing the metric based on general relativity (or the Einstein-Hilbert action) is perturbatively nonrenormalizable, requiring an infinite number of counterterms and leading to a loss of predictability at the quantum level. In terms of power counting in quantum field theory, this nonrenormalizability arises because the Newton coupling $G_\tx{N}$ (or equivalently the inverse of the Planck mass squared) acts as a perturbation parameter with a negative mass dimension.

A further complication arises from the treatment of the metric field itself. To analyze gravitational fluctuations, the metric $g_{\mu\nu}$ is typically expanded around a fixed background metric $\bar g_{\mu\nu}$ as $g_{\mu\nu} = \bar g_{\mu\nu} + \mf h_{\mu\nu}$, where $\mf h_{\mu\nu}$ represents the fluctuation field. The inverse metric $g^{\mu\nu}$, which is needed to evaluate the Ricci scalar curvature, contains an infinite series of $\mf h_{\mu\nu}$. This expansion introduces an infinite number of interaction terms involving the fluctuation field, while the Newton coupling $G_\tx{N}$ is insufficient to cancel the ultraviolet divergences arising from quantum loops with arbitrarily many vertices. Together, these issues highlight the need for a novel approach to address the origin of spacetime and mass scales.

In this work, we explore a scale-invariant model for gravity and matter fields under the irreversible vierbein postulate~\cite{Matsuzaki:2020qzf,Maitiniyazi:2023hts}, which preserves local Lorentz (LL) gauge symmetry and general-coordinate (GC) invariance at a specific energy scale $\Lambda_\text{G}$. Our approach employs the degenerate limit~\cite{Tseytlin:1981ks,Horowitz:1990qb,Floreanini:1991cw} and classical scale invariance~\cite{Wetterich:1983bi,Bardeen:1995kv}, allowing for degenerate background spacetimes and forbidding dimensionful parameters, such as the Planck mass, at $\Lambda_\text{G}$. Although classical scale invariance is sometimes criticized for lacking a quantum foundation, it can alternatively be interpreted as one of the critical points within the multicritical point principle, as emphasized in Refs.~\cite{Hamada:2021jls,Kawai:2021lam,Hamada:2022soj,Kawai:2023viy}, or as self-tuned criticality in asymptotically safe gravity~\cite{Wetterich:2016uxm}.

The model simultaneously realizes scalegenesis and pregeometry: the Planck mass arises from the vacuum expectation value $\langle\phi\rangle$ of a scalar field, while a background spacetime emerges from $\langle e^\ba{}_\mu\rangle$ in the vierbein dynamics. For related models, see Refs.~\cite{Kubo:2018kho,Volovik:2023pcm,Edery:2015wha,Edery:2019txq,Edery:2014nha,Ghilencea:2018dqd,Ghilencea:2018thl,Shtanov:2023lci,Karananas:2024xja} (Planck-scale generation) and Refs.~\cite{Akama:1978pg,Terazawa:1980vf,Akama:1981dk,Akama:1979tm,Akama:1981kq,Akama:1990ur,Akama:1991pv,Akama:1991np,Akama:1994ehj,Wetterich:2021ywr,Wetterich:2021cyp,Wetterich:2021hru,Floreanini:1990cf,Volovik:2021wut} (pregeometry). To achieve this, we derive the effective potential as a function of $\phi$ and $e^\ba{}_\mu$ and demonstrate the existence of vacuum expectation values.

This paper is organized as follows: In Sec.~\ref{sec: Setup} we provide a brief overview of LL and GC symmetries, along with the degenerate limit and the generation of spacetime, forming the basis for our discussion. Further, we introduce the irreversible vierbein postulate and the scale-invariant action, establishing the foundational principles of our model. In Sec.~\ref{sec: Spacetime and Planck mass}, we present our main result: the simultaneous generation of a nonvanishing flat background field and the Planck scale, driven by quantum effects of fermionic degrees of freedom. In Sec.~\ref{sec:Conclusions}, we conclude and discuss our results.

\section{Setup}
\label{sec: Setup}

In this section, we summarize the setup for our gravitational model and review results on spacetime generation from earlier studies. The construction of the gravitational action is based on the following working assumptions:
\begin{enumerate}
\item Gravitational interactions are invariant under $SO(1,3)_\text{LL}\times \text{GC}$~\cite{Matsuzaki:2020qzf,Maitiniyazi:2023hts};
\item Irreversible vierbein postulate: the action allows for the degenerate limit~\cite{Tseytlin:1981ks,Horowitz:1990qb,Floreanini:1991cw};
\item Classical scale invariance~\cite{Wetterich:1983bi,Bardeen:1995kv}.
\end{enumerate}

Below, we elaborate on these assumptions, explaining their definitions and the motivations for adopting them. A detailed theoretical description of gravitational theories with $SO(1,3)_\text{LL}\times \text{GC}$ invariance can be found in Ref.~\cite{Maitiniyazi:2023hts}.

\subsection{LL and GC symmetries}

We consider a gravitational theory invariant under $SO(1,3)_\text{LL}\times \text{GC}$ transformation. 
As the gravitational degrees of freedom, we introduce a vierbein $e^{\ba}{}_\mu(x)$, which is essential for defining spinor fields on curved spacetimes. Here and throughout, greek indices $\mu$, $\nu$, $\cdots$ represent spacetime indices, while bold Roman indices $\ba$, $\bb$, $\cdots$ denote LL indices in the fundamental representation. The vierbein transforms under LL and GC transformations as
\al{
e^\ba{}_{\mu}(x) &\os{\text{LL}}{\to} e'^\ba{}_{\mu}(x)
    = \Lambda^{\ba}{}_\bb(x) e^\bb{}_{\mu}(x), \\
e^\ba{}_{\mu}(x) &\os{\text{GC}}{\to} e'^\ba{}_{\mu}(x')
    = e^\ba{}_{\nu}(x) [M^{-1}(x)]^\nu{}_\mu,
}
where $\Lambda^{\ba}{}_\bb = (e^{\theta(x)})^\ba{}_\bb$ and $[M^{-1}(x)]^\nu{}_\mu = \frac{\p x^\nu}{\p x'^\mu}$ are the matrix elements of the LL gauge and GC transformations, respectively. The metric is constructed from the vierbein as $g_{\mu\nu}(x) = \eta_{\ba\bb}e^\ba{}_\mu(x) e^\bb{}_\nu(x)$, where $\eta_{\ba\bb} = \diag(-1,1,1,1) = \eta^{\ba\bb}$ is the tangent-space metric, and $\eta^{\ba\bb}$ is its inverse.

Additionally, $[M^{-1}(x)]^\mu{}_\nu$ in the GC transformation satisfies 
\al{
\p_{[\lambda}(M^{-1})^\mu{}_{\nu]}(x) = 0,
\label{eq: GC condition}
}
which, in four-dimensional spacetimes, fixes 24 degrees of freedom.

The GC gauge field $\Upsilon^\alpha{}_{\beta\mu}(x)$ transforms under the GC transformation as
\al{
\Upsilon^\alpha{}_{\beta\mu}\fn{x} \sr{\tx{GC}}{\to} \Upsilon^{\pr\alpha}{}_{\beta\mu}\fn{x'} = \sqbr{M^\alpha{}_\gamma\fn{x}\Upsilon^\gamma{}_{\delta\nu}\fn{x}\pa{M^{-1}}^\delta{}_\beta\fn{x} - \p_\nu M^\alpha{}_\gamma\fn{x}\pa{M^{-1}}^\gamma{}_\beta\fn{x}}\pa{M^{-1}}^\nu{}_\mu\fn{x}.
}
It is decomposed into symmetric and antisymmetric parts as
\al{
\Upsilon^\alpha{}_{\beta\mu}\fn{x} = \Upsilon^\alpha{}_{(\beta\mu)}\fn{x} + \Upsilon^\alpha{}_{[\beta\mu]}\fn{x},
}
with the symmetric and antisymmetric parts defined as
\al{
\Upsilon^\alpha{}_{(\beta\mu)} &= \frac{1}{2} \pn{\Upsilon^\alpha{}_{\beta\mu} + \Upsilon^\alpha{}_{\mu\beta}}, &
\Upsilon^\alpha{}_{[\beta\mu]} &= \frac{1}{2} \pn{\Upsilon^\alpha{}_{\beta\mu} - \Upsilon^\alpha{}_{\mu\beta}}.
}
The antisymmetric part, corresponding to the (con)torsion, has 24 degrees of freedom, which are fixed by the condition~\eqref{eq: GC condition} and do not contribute to the dynamics. Thus, the GC gauge field is identified with the symmetric part $\Upsilon^\alpha{}_{(\beta\mu)}\fn{x}$, which has 40 degrees of freedom. If the condition~\eqref{eq: GC condition} on the transformation matrix $M$ were not imposed, the corresponding symmetry would enlarge to $GL(4)$ rather than GC.\footnote{In this case, $[M^{-1}(x)]^\mu{}_\nu$ is regarded as a group element of $GL(4)$ gravity~\cite{Ohanian:1969xhl,Nakanishi:1979fg,Percacci:1990wy,Floreanini:1991gi,Floreanini:1991cw,Floreanini:1994ypa,Floreanini:1995ie,Percacci:2009ij,Tomboulis:2011qh}, which has 64 degrees of freedom.}

If we impose the metricity condition on $\Upsilon^\alpha{}_{(\beta\mu)}\fn{x}$, i.e.,
\al{
\os{\Upsilon}{\nabla}_\alpha g_{\beta\mu}(x) = \p_\alpha g_{\beta\mu}(x) - \Upsilon^\lambda{}_{(\alpha\beta)}\fn{x}g_{\lambda\mu}(x) = 0,
}
then the GC gauge field reduces to the Levi-Civita connection as its solution:
\al{
\Upsilon^\alpha{}_{(\beta\mu)}\fn{x}
= \os{g}\Gamma^\alpha{}_{\beta\mu}(x) 
= \frac{g^{\alpha\gamma}(x)}{2} \pn{\p_\beta g_{\mu\gamma}(x) + \p_\mu g_{\beta\gamma}(x) - \p_\gamma g_{\beta\mu}(x)}.
}

The LL gauge field is defined on the vector representation as $\omega_\mu\fn{x}={1\ov2}\omega_{\bc\bd\mu}\fn{x}T^{\bc\bd}$, or more explicitly as $\omega^\ba{}_{\bb\mu}\fn{x}={1\ov2}\omega_{\bc\bd\mu}\fn{x}\pn{T^{\bc\bd}}{}^\ba{}_\bb$, where $\pn{T^{\bc\bd}}{}^\ba{}_\bb=\eta^{\bc\ba}\delta^\bd_\bb-\eta^{\bd\ba}\delta^\bc_\bb$ are the $SO(1,3)$ generators. Similarly, on the spinor representation, it is given by $\omega_\mu\fn{x}={1\ov2}\omega_{\bc\bd\mu}\fn{x}\sigma^{\bc\bd}$, with $\sigma^{\ba\bb} = \frac{1}{4}[\gamma^\ba, \gamma^\bb]$.
Note that $\omega^\ba{}_{\bb\mu} = -\omega_\bb{}^\ba{}_{\mu}$, and the Dirac matrices $\gamma^\ba$ satisfy the Clifford algebra $\{\gamma^\ba, \gamma^\bb\} = 2\eta^{\ba\bb}$. The field strength of $\omega^\ba{}_{\bb\mu}$ is given by
\al{
\os{\omega}{{\mathcal F}}_{\mu\nu} &= \p_\mu \omega_\nu - \p_\nu \omega_\mu - [\omega_\mu, \omega_\nu], &
\os{\omega}{{\mathcal F}}^{\ba\bb}{}_{\mu\nu} &= \p_\mu \omega^{\ba\bb}{}_\nu - \p_\nu \omega^{\ba\bb}{}_\mu - [\omega_\mu, \omega_\nu]^{\ba\bb},
}
where $\pn{\omega_\mu \omega_\nu}^{\ba\bb} := \omega^\ba{}_{\bc\mu} \omega^{\bc\bb}{}_\nu$, etc., on the vector representation.

The LL gauge field transforms under GC and LL transformations as follows:
\al{
\omega^\ba{}_\bb{}_\mu(x) &\os{\text{GC}}{\to} \omega'^\ba{}_\bb{}_\mu(x') = \omega^\ba{}_{\bb\nu}(x) [M^{-1}(x)]^\nu{}_\mu, \\
\omega^\ba{}_\bb{}_\mu(x) &\os{\text{LL}}{\to} \omega'^\ba{}_\bb{}_\mu(x) = \Lambda^\ba{}_\bc(x) \omega^\bc{}_{\bd \mu}(x) (\Lambda^{-1})^\bd{}_\bb(x) - \p_\mu \Lambda^\ba{}_\bc(x) (\Lambda^{-1})^\bc{}_\bb(x).
}

A spinor field transforms as a scalar under the GC transformation,
\al{
\psi \os{\text{GC}}{\to} \psi,
}
while under the LL transformation, it transforms as a spinor,
\al{
\psi \os{\text{LL}}{\to} S\fn{\Lambda} \psi,
}
where $S\fn{\Lambda}$ represents the spinor representation of the LL transformation $\Lambda$.

\subsection{Degenerate limit and spacetime generation}

In this subsection, we explain the concept of the degenerate limit~\cite{Tseytlin:1981ks,Horowitz:1990qb}, which necessarily arises when spacetime involves topology changes. The degenerate limit refers to a situation where some eigenvalues of the vierbein vanish, resulting in a zero determinant for the vierbein:
\al{
\ab{e\fn{x}} := \det_{\ba,\mu} e^{\bf a}{}_\mu = 0.
\label{eq: degenerate limit}
}
We assume that the gravitational action remains well defined and free of divergent terms in the degenerate limit. This assumption is what we call the ``irreversible vierbein postulate." A similar principle is applied in the Standard Model, where the action is constructed to be well defined in the weak field limit \(H \to 0\) by excluding terms like \(1/H^\dagger H\). Such terms would introduce singularities as the Higgs field approaches zero, rendering the action ill defined. Analogously, the irreversible vierbein postulate ensures that the gravitational action avoids such singularities in the degenerate limit, allowing the theory to remain meaningful even in these extreme configurations.

To clarify the statement, let us consider an example. The kinetic term of spinor fields,
\al{
-\ab{e} \overline\psi\fn{x}e_{\bf a}{}^\mu\fn{x} \gamma^{\bf a} \p_\mu \psi\fn{x},
}
contains the combination of vierbeins $\ab{e}e_{\bf a}{}^\mu$, which is compatible with the degenerate limit. For a diagonalized vierbein in four-dimensional spacetime, $e^\ba{}_\mu = \diag(\lambda_1, \lambda_2, \lambda_3, \lambda_4)$, with eigenvalues $\lambda_i$ and its inverse $e_\ba{}^\mu = \diag(\lambda_1^{-1}, \lambda_2^{-1}, \lambda_3^{-1}, \lambda_4^{-1})$, the condition~\eqref{eq: degenerate limit} is realized when one or more eigenvalues vanish, $\lambda_i \to 0$, since $\ab{e} = \lambda_1 \lambda_2 \lambda_3 \lambda_4$. For the spinor kinetic term, we compute the following:
\al{
\ab{e}e_{\bf a}{}^\mu &= \lambda_1\lambda_2\lambda_3\lambda_4  \diag(\lambda_1^{-1},\,\lambda_2^{-1},\,\lambda_3^{-1},\,\lambda_4^{-1})
= \diag(\lambda_2\lambda_3\lambda_4,\,\lambda_1\lambda_3\lambda_4,\,\lambda_1\lambda_2\lambda_4,\,\lambda_1\lambda_2\lambda_3),
}
where none of the elements diverge in the degenerate limit $\lambda_i \to 0$. Hence, the kinetic term for spinor fields is compatible with the degenerate limit. 

In contrast, the combination $\ab{e}g^{\mu\nu} = \ab{e} e^\mu{}_{(\ba} e^\nu{}_{\bb)} \eta^{\ba\bb}$ becomes divergent in the degenerate limit. Each component contains an inverse eigenvalue:
\al{
\ab{e} g^{\mu\nu} &= \lambda_1\lambda_2\lambda_3\lambda_4 \diag(\lambda_1^{-2},\,\lambda_2^{-2},\,\lambda_3^{-2},\,\lambda_4^{-2}) \nn
&= \diag( \lambda_1^{-1}\lambda_2\lambda_3\lambda_4,\,\lambda_1\lambda_2^{-1}\lambda_3\lambda_4,\,\lambda_1\lambda_2\lambda_3^{-1}\lambda_4,\,\lambda_1\lambda_2\lambda_3\lambda_4^{-1}),
}
demonstrating the divergence. As a result, the kinetic term of a scalar field, $-\ab{e} g^{\mu\nu} \p_\mu \phi \p_\nu \phi$, diverges in the degenerate limit and is therefore forbidden. Similarly, requiring the action to remain finite in the degenerate limit also prohibits the kinetic term of a gauge field. For further details on the degenerate limit, see Appendix~C of Ref.~\cite{Maitiniyazi:2023hts}.

We now explain why the action in the degenerate limit is considered. In the standard procedure for computations in gravitational theories, one typically assumes an expansion of the gravitational fields, such as $g_{\mu\nu} = \bar g_{\mu\nu} + \mf h_{\mu\nu}$ in the Einstein-Hilbert theory or $e^\ba{}_\mu = \bar e^\ba{}_\mu + \mathfrak{e}^\ba{}_\mu$ in the Einstein-Cartan theory. Here, $\mf h_{\mu\nu}$ and $\mathfrak{e}^\ba{}_\mu$ are fluctuation fields, while $\bar g_{\mu\nu}$ and $\bar e^\ba{}_\mu$ are nondegenerate background fields with inverses $\bar g^{\mu\nu}$ and $\bar e_\ba{}^\mu$. Such background fields satisfy $g_{\mu\rho} g^{\rho\nu} = \delta_\mu^\nu$ and $e^\ba{}_\mu e_\ba{}^\nu = \delta_\mu^\nu$ (or $e^\ba{}_\mu e_\bb{}^\mu = \delta_\bb^\ba$), allowing standard computations. However, these expansions introduce an infinite number of fluctuation fields. For example, the inverse metric field expands as
\al{
g^{\mu\nu} = \bar g^{\mu\nu} - \mf h^{\mu\nu} + \mf h^\mu{}_{\rho}\mf h^{\rho \nu} + \cdots,
\label{eq: inverse metric expansion}
}
where indices are raised and lowered by the background metric. This expansion implies that terms involving the inverse metric field $g^{\mu\nu}$ generate a finite number of vertices even at the classical level, resulting in a nonlinear structure that is nonrenormalizable, as occurs in the Einstein-Hilbert action.

This situation is similar to the $O(N)$ nonlinear sigma model, where the fields $\vec\phi = (\pi^1, \cdots, \pi^{N-1}, \sigma)$ satisfy the constraint $\varphi^i \varphi^i = 1$, with $\varphi^i = \phi^i / f_\pi$ and $\langle \phi^N \rangle = \langle \sigma \rangle = f_\pi$. This constraint makes the theory highly nonlinear and perturbatively nonrenormalizable, even if its action appears simple. From the perspective of the $O(N)$ linear sigma model, which is perturbatively renormalizable, the nonlinearity arises from expanding fields around a fixed symmetry-broken vacuum $\langle \varphi^i \rangle = \delta^{iN}$ as $\sigma = \sqrt{f_\pi^2 - (\pi^i)^2}$, leading to a nonlinear action for $\pi^i$. In gravitational theories, assuming a background vierbein $\bar e^\ba{}_\mu$ (or metric $\bar g_{\mu\nu}$) is analogous to assuming a finite vacuum \( f_\pi \) in the sigma model. However, such a finite vacuum \( f_\pi \) should be dynamically generated by the linear sigma model.

The degenerate limit instead defines a symmetric vacuum $\langle e^\ba{}_\mu \rangle = \bar e^\ba{}_\mu = 0$, resolving the nonlinearity and establishing a ``linear" theory of gravity. In Ref.~\cite{Maitiniyazi:2023hts}, we demonstrated how the dynamics of the action with the degenerate limit generate a background vierbein field. In the next subsection, we summarize the status of this approach.

\subsection{Effective potential for vierbein}

The irreversible vierbein postulate prevents the introduction of {\it a priori} nondegenerate background fields for gravitational fields. This means that field expansions around a fixed background, such as Eq.~\eqref{eq: inverse metric expansion}, cannot be employed in the classical action. In the degenerate limit, only a limited number of terms can contribute to the action. Notably, the kinetic terms for scalar and gauge fields are prohibited, whereas spinor fields retain their kinetic term and remain dynamical at the classical level, suggesting that spinor fields might serve as the dynamical origin of spacetime and particles.

In previous works~\cite{Matsuzaki:2020qzf,Maitiniyazi:2023hts}, we formulated an action accommodating the degenerate limit. At a certain ultraviolet scale $\Lambda_\text{G}$, the action is given by:
\al{
S &= \int\df^4x\ab{e\fn{x}}\Bigg[-\Lambda_\text{cc} -
\overline\psi\fn{x}\Big(e_{\bf a}{}^\mu\fn{x} \gamma^{\bf a}  \mathcal D_\mu + m_f \Big)\psi\fn{x} 
+ \frac{M_\text{pl}^2}{2}e_{[\bf a}{}^\mu\fn{x}e_{\bf b]}{}^\nu\fn{x}F^{\bf ab}{}_{\mu\nu}\fn{x}\Bigg],
\label{eq: previous action}
}
where $\Lambda_\text{cc}$ is the cosmological constant, and the (reduced) Planck mass squared $M_\text{pl}^2$ is related to the Newton constant $G_\tx{N}$ as $M_\text{pl}^2 = 1/8\pi G_\tx{N}$.\footnote{
These constants can depend on combinations of fields that are scalar under GC and LL transformations.
} The covariant derivative acting on the spinor field $\psi$ is defined as $\mathcal D_\mu = \p_\mu + \frac{1}{2}\omega_{{\bf bc}\mu}\fn{x}\sigma^{\bf bc}$. Importantly, the irreversible vierbein postulate prohibits the kinetic term of $\omega^{\ba\bb}{}_\mu$, which is a square of its field strength $F^{\ba\bb}{}_{\mu\nu}$, but allows the invariant term linear in $F^{\ba\bb}{}_{\mu\nu}$. This linear term is a unique feature of the vierbein, which transforms as a fundamental representation under the LL gauge transformation while being a spacetime vector field—distinct from ordinary Yang-Mills gauge theory.

\begin{figure}
\centering
\includegraphics[width=0.5\linewidth]{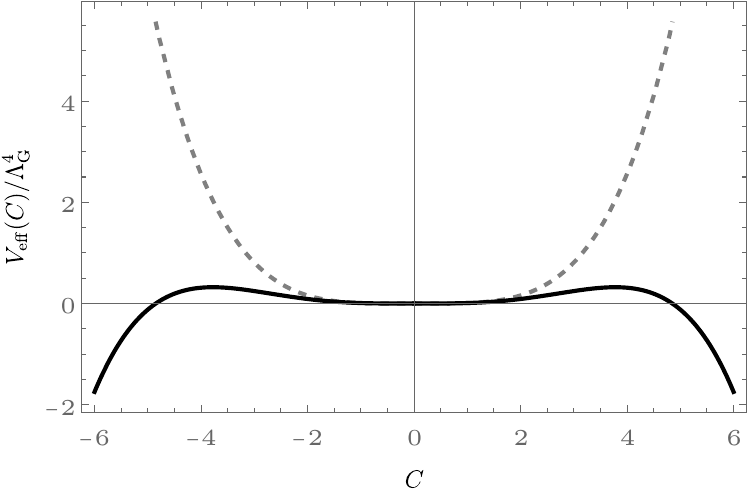}
\caption{
Shape of the effective potential~\eqref{eq: effective potential} with \(\Lambda_{\rm cc} = 0.01\) and \(m_f/\Lambda_\tx{G} = 1\) (black solid line). The dashed gray line shows the tree-level potential \( V_\text{tree}(C) = \Lambda_{\rm cc} C^4 \).
\label{fig:effective potential}
}
\end{figure}

Making a simple ansatz for the background vierbein field as \( e^\ba{}_\mu = C \delta^\ba_\mu \), corresponding to setting \( \lambda_1 = \lambda_2 = \lambda_3 = \lambda_4 = C \), we derive the effective potential for \( C \). This treatment is analogous to the minisuperspace model~\cite{Hartle:1983ai,Halliwell:1988ik,Vilenkin:1994rn}. 
At the classical level, the potential \( V_\text{eff}(C) = \Lambda_\text{cc} C^4 \) yields the trivial vacuum \(\langle C \rangle = 0\) for \(\Lambda_\text{cc} > 0\), respecting the degenerate limit \( C \to 0 \). A nontrivial vacuum arises when spinor loop effects are included. The quantum dynamics of the spinor field deform the effective potential, leading to \(\langle e^\ba{}_\mu \rangle \neq 0\). Specifically, by considering the bare perturbation, we obtain:
\al{
V_\text{eff}(C) = \Lambda_\text{cc} C^4 - \frac{m_f^4 C^4}{16\pi^2}\log\left(\frac{m_f^2 C^2}{\Lambda_\text{G}^2}\right),
\label{eq: effective potential}
}
where the second term is generated by the spinor one-loop effect, taking the form of a Coleman-Weinberg potential~\cite{Coleman:1973jx,Floreanini:1991gi}. The shape of the effective potential~\eqref{eq: effective potential} is shown in Fig.~\ref{fig:effective potential} for reference values \(\Lambda_\text{cc} = 0.01\) and \( m_f/\Lambda_\tx{G} = 1 \). 

The potential takes a runaway form and does not exhibit a stable vacuum. This instability might be corrected by the effects of higher-dimensional operators~\cite{Maitiniyazi:2023hts}, or the runaway behavior could signify spacetime expansion. In either case, the symmetric vacuum is metastable, allowing for a nonzero \( C \).

Regarding GC invariance, the effective potential~\eqref{eq: effective potential} was derived using a cutoff based on \(\eta^{\mu\nu}p_\mu p_\nu = p^2 < \Lambda_\text{G}^2\). Alternatively, applying the cutoff \( g^{\mu\nu}p_\mu p_\nu = C^{-2}\eta^{\mu\nu}p_\mu p_\nu = C^{-2}p^2 < \Lambda_\text{G}^2 \), the effective potential becomes~\cite{Floreanini:1993na,Oda:2024nda}
\al{
V_\text{eff}(C) = \left[ \Lambda_\text{cc} - \frac{m_f^4}{16\pi^2}\log\left(\frac{m_f^2}{\Lambda_\text{G}^2}\right) \right] C^4.
\label{eq: effective potential other}
}
Here, no Coleman-Weinberg type term \(\sim C^4\log C\) appears. From the perspective of GC invariance, Eq.~\eqref{eq: effective potential other} may always be preferred. When the overall coefficient of \( C^4 \) becomes negative, the symmetric vacuum \(\langle C \rangle = 0\) becomes unstable, leading to spacetime generation.

As discussed in Ref.~\cite{Floreanini:1993na}, the form of the effective potential depends on the interpretation of the base metric. If the flat metric \(\eta_{\ba\bb}\) is interpreted as the base metric,\footnote{
Note, however, that this implicit interpretation, as taken in Ref.~\cite{Maitiniyazi:2023hts}, can only realize flat spacetime.
}
then the effective potential takes the Coleman-Weinberg form~\eqref{eq: effective potential}. Conversely, if the base metric is \(g_{\mu\nu} = C^2 \delta_\mu^\ba \delta_\nu^\bb \eta_{\ba\bb}\), the effective potential adopts the \(C^4\) form~\eqref{eq: effective potential other}. In our previous work~\cite{Maitiniyazi:2023hts}, the former interpretation was implicitly assumed to discuss the generation of nontrivial \( C \). In this work, we explore the latter possibility.

It is worth emphasizing that the irreversible vierbein postulate enables the limit \( C \to 0 \) to be taken continuously. With this postulate, the classical energy (and action) is determined to vanish at \( C = 0 \), establishing a continuous connection between the configurations \( C = 0 \) and \( C \neq 0 \). Notably, \( C = 0 \) does not correspond to a distinct phase: It cannot serve even as a metastable local minimum for the parameter choice specified below. Instead, it appears as a local maximum, as demonstrated in Eqs.~\eqref{eq: effective potential} and \eqref{eq: effective potential other}. In summary, the irreversible vierbein postulate assumes the \emph{allowance} of degeneracy, which consequently forbids any scalar or ordinary vector kinetic terms at the scale \( \Lambda_\text{G} \).

\subsection{Irreversible vierbein postulated action with scale invariance and aiming scenario}
\label{sec: IRREVERSIBLE}

In our previous work~\cite{Maitiniyazi:2023hts}, we demonstrated that the terms in the action~\eqref{eq: previous action} are heavily constrained by the degenerate limit. However, dimensionful parameters, such as the cosmological constant, the Planck mass, and the spinor mass, were introduced independently without specifying their origins. Notably, the spinor mass plays a crucial role in achieving a nontrivial vacuum \(\langle C \rangle \neq 0\) through the effective potentials~\eqref{eq: effective potential} or~\eqref{eq: effective potential other}. This approach leaves the origin of background spacetimes dependent on the introduction of such an intrinsic scale, while the Planck scale remains entirely disconnected from spacetime generation. In this work, we aim to address the origin of the spinor mass and investigate whether it is related to the Planck mass.

To avoid introducing intrinsic mass scales, we impose (classical) scale invariance~\cite{Wetterich:1983bi,Bardeen:1995kv} on the action and introduce a real singlet scalar field \(\phi\) coupled to both the spinor and gravitational fields. Specifically, the action is formulated based on the irreversible vierbein postulate and scale invariance as
\begin{align}
S &= \int\df^4x\ab{e\fn{x}}\Bigg[
-\overline\psi\fn{x}\Big(e_{\bf a}{}^\mu\fn{x} \gamma^{\bf a}  \mathcal D_\mu + y\phi\fn{x} \Big)\psi\fn{x}
+ \frac{\xi \phi^2\fn{x}}{2}e_{[\bf a}{}^\mu\fn{x}e_{\bf b]}{}^\nu\fn{x}F^{\bf ab}{}_{\mu\nu}\fn{x}
- \frac{\lambda}{4!}\phi^4\fn{x}
\Bigg],
\label{eq: proposed action}
\end{align}
where all parameters (\(y\), \(\xi\), and \(\lambda\)) are dimensionless. The degenerate limit prohibits kinetic terms for both the scalar and LL gauge fields.

In this action, we aim to simultaneously explain the origins of background spacetimes and the Planck scale through the dynamics of spinor field fluctuations. Quantum effects of the spinor fields can induce a vacuum with \(\langle e^\ba{}_\mu \rangle \neq 0\) and \(\langle \phi \rangle \neq 0\). In such a vacuum, the Yukawa term \(y\phi\overline\psi \psi\) generates the spinor mass \(m_f = y\langle \phi \rangle\), while the Planck mass squared originates from the nonminimal coupling as \(M_\text{pl}^2 = \xi \langle \phi \rangle^2\).

\section{Spacetime and Planck mass generation}
\label{sec: Spacetime and Planck mass}

In this work, we aim to simultaneously generate the Planck mass scale and a background spacetime within the framework of the action~\eqref{eq: proposed action}. At this stage, only the spinor field is dynamical, while the other fields become dynamical indirectly through spinor field loops. Consequently, the dynamics of the other fields are subleading. Thus, at the one-loop level, we can neglect the quantum effects of \(\phi\), \(e^\ba{}_\mu\), and \(\omega^\ba{}_\bb{}_\mu\).

\subsection{Action in conformal coordinates}

To derive the effective potential for \(\phi\) and \(e^\ba{}_\mu\), we assume the background fields take the following form:
\al{
&e^\ba{}_\mu(t,\bs x) = \bar e^\ba{}_\mu(t) = \diag \left(1,\bar C(t),\bar C(t),\bar C(t) \right),&
&\phi(t,\bs x) = \bar\phi(t),
}
where \(\bar C(t)\) corresponds to the scale factor. Introducing the conformal coordinates \(x = (\eta, \bs x)\) with \(\df t = \bar C(\eta) \df\eta\), the background metric becomes
\al{
{\df s}^2
    &= \bar g_{\mu\nu}\fn{x}\df x^\mu \df x^\nu
    = \bar C^2\fn{\eta}\pn{-\df\eta^2 + \df\bs x^2}
    = \bar C^2\fn{\eta} \eta_{\mu\nu}\df x^\mu \df x^\nu,
}
and the corresponding vierbein is
\al{
\bar e^\ba{}_\mu\fn{x} = \bar C\fn{\eta} \delta^\ba_\mu.
}

In conformal coordinates, the terms in the action~\eqref{eq: proposed action} become
\begin{align}
&\ab{\bar e} = \bar C^4(\eta),&
\ab{\bar e} \bar e_{[\ba}{}^\mu \bar e_{\bb]}{}^\nu = \bar C^2(\eta) \delta_{[\ba}^\mu \delta^\nu_{\bb]},&
&\bar e_{[\ba}{}^\mu \bar e_{\bb]}{}^\nu \bar F^{\ba\bb}{}_{\mu\nu} = \bar R,
\label{eq: absolute BF vierbein}
\end{align}
where \(\bar F^{\ba\bb}{}_{\mu\nu}\) is determined by the background LL gauge field \(\bar\omega^\ba{}_\bb{}_\mu\), obtained as a solution to the equations of motion, and \(\bar R\) is the Ricci scalar. Notably, at the degenerate limit (\(\bar C = 0\)), we have \(\bar \omega^{\ba\bb}{}_\mu = 0\) and \(\bar R = 0\).

The action~\eqref{eq: proposed action} in conformal coordinates thus reads
\begin{align}
S &= \int\df\eta\,\df^3\bs x\,\bar C^4\fn{\eta} \Bigg[
-\overline\psi\fn{x}\Big(\Slash{\bar{\mathcal{D}}} + y\bar{\phi}\fn{\eta} \Big)\psi\fn{x}
+ \frac{\xi \bar R}{2} \bar\phi^2\fn{\eta}
- \frac{\lambda}{4!}\bar\phi^4\fn{\eta} \Bigg],
\label{eq: proposed action in De Sitter background}
\end{align}
where \(\Slash{\bar{\mathcal{D}}} = \bar e_\ba{}^\mu \gamma^\ba \bar{\mathcal{D}}_\mu = \bar C^{-1} \delta_\ba^\mu \gamma^\ba \bar{\mathcal{D}}_\mu\).

\subsection{Effective potential for vierbein and scalar fields}

In the classical action~\eqref{eq: proposed action in De Sitter background}, the tree-level potential for \(\bar C\) and \(\bar\phi\) is given by:
\al{
V_\text{tree}\fn{\bar\phi,\bar C} &= \left( -\frac{\xi \bar R}{2} \bar\phi^2 + \frac{\lambda}{4!} \bar\phi^4 \right) \bar C^4.
\label{eq:tree level potential}
}
Here, we assume a constant scalar curvature~\cite{Giddings:1991qi}. For \(m_\phi^2 := \xi \bar R > 0\), the classical potential has a stable vacuum at \(\langle \bar\phi \rangle = \sqrt{6m_\phi^2 / \lambda} \neq 0\), which gives rise to the Planck mass \(M_\tx{pl} = \xi \langle \bar\phi \rangle\).

At this vacuum, the cosmological constant is \(\Lambda_\tx{cc} \equiv -\frac{\xi \bar R}{2} \langle \bar\phi \rangle^2 + \frac{\lambda}{4!} \langle \bar\phi \rangle^4 = -3m_\phi^4 / 2\lambda < 0\). Thus, the potential near the symmetric vacuum \(\langle \bar C \rangle = 0\) is unstable, leading to spacetime generation.

In contrast, for \(m_\phi^2 < 0\), the potential supports only the symmetric vacuum \(\langle \bar\phi \rangle = \langle \bar C \rangle = 0\). Therefore, the case \(m_\phi^2 > 0\) is necessary for realizing scalegenesis and pregeometry, even at the tree level in our model.

To analyze the stability of the vacuum, we consider the spinor one-loop effect. Integrating out the spinor fields under fixed background fields $\bar C$ and $\bar\phi$, the one-loop potential is
\al{
V_\text{1-loop}(\bar\phi,\bar C)
    &= -\Tr\ln\fn{-\Slash{\mathcal D} - y\bar\phi},
}
where $\bar C$ is included implicitly through the volume element. On a curved background, we evaluate this loop kernel using the heat kernel expansion, which expresses the one-loop term in powers of curvature operators and their derivatives. Specifically, 
\al{
\Tr\ln\fn{-\Slash{\mathcal D} - y\bar\phi}
&= \frac{1}{2} \int_0^\infty \df z \ln(z + m_f^2) \frac{1}{2\pi i} \int_{\gamma-i\infty}^{\gamma+i\infty} \df s \, e^{sz} \Tr e^{-s\bar\Delta},
}
where $\bar\Delta = -\Slash{\bar{\mathcal D}}^2$, $m_f = y\bar\phi$, and the inverse Laplace transformation (Merlin formula) is used for the delta function~\cite{Wetterich:2019zdo}. The term $K(s) \equiv \Tr e^{-s\bar\Delta}$, known as the heat kernel, satisfies the thermal diffusion equation and is expanded as
\al{
K(s) = \Tr e^{-s\bar\Delta} 
= \frac{1}{(4\pi)^2} \int \df^4x \sqrt{-\bar g} \left( a_0 s^{-2} + a_2 s^{-1} + O(s^0) \right).
}
Here, $\sqrt{-\bar g} = |\bar e| = \bar C^4$, and $a_i$ are heat kernel coefficients depending on the field representation on which $\bar\Delta$ acts. For the Dirac spinor field, the first two coefficients are $a_0 = 4$ and $a_2 = 4\bar R/3$~\cite{Vassilevich:2003xt}.

Including the tree-level potential~\eqref{eq:tree level potential}, the regularized effective potential is given by
\al{
V_{\rm eff}(\bar{\phi},\bar{C})
&= V_\text{tree}(\bar\phi,\bar C) + V_\text{1-loop}(\bar\phi,\bar C)\nn
&= \left[ -\frac{\xi\bar R}{2} \bar{\phi}^2 + \frac{\lambda}{4!}\bar{\phi}^4 - \frac{(y\bar{\phi})^2}{16\pi^2}\left( (y\bar{\phi})^2 - \frac{2}{3}\bar R \right) \log\fn{\frac{(y\bar{\phi})^2}{\Lambda_\text{G}^2}} \right]\bar{C}^4.
\label{eq:effective potential}
}
For $\bar R = 0$, this potential reproduces Eq.~\eqref{eq: effective potential other} with $m_f = y\bar\phi$ and $\Lambda_\tx{cc} = \frac{\lambda}{4!}\bar\phi^4$. Thus, the background curvature scalar provides the origin of scalar and fermion masses via spontaneous symmetry breaking. 

Similar mass generation scenarios involving a Weyl-invariant gravitational sector with scalar curvature, as well as gravitational-Higgs portal couplings to the Standard Model, have been explored in the literature~\cite{Kubo:2018kho,Volovik:2023pcm,Edery:2015wha,Edery:2019txq,Edery:2014nha,Ghilencea:2018dqd,Ghilencea:2018thl,Shtanov:2023lci,Karananas:2024xja}. However, these approaches do not incorporate the notions of the degenerate limit or pregeometry.

\begin{figure}
\centering
\includegraphics[width=0.8\linewidth]{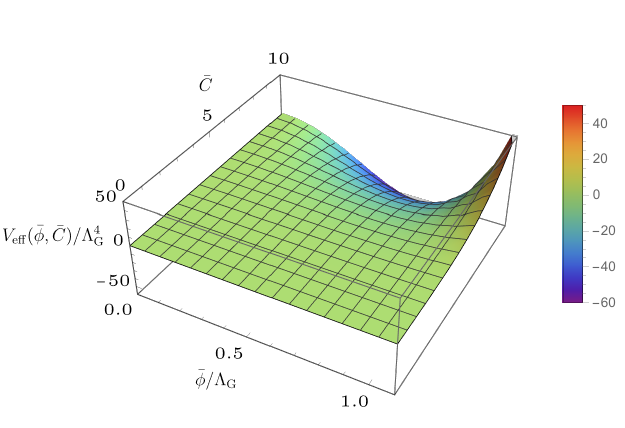}
\includegraphics[width=0.45\linewidth]{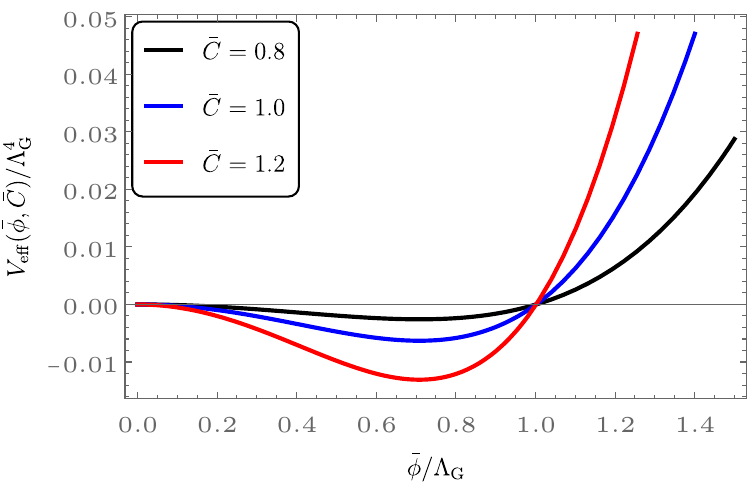}
\hspace{5ex}
\includegraphics[width=0.45\linewidth]{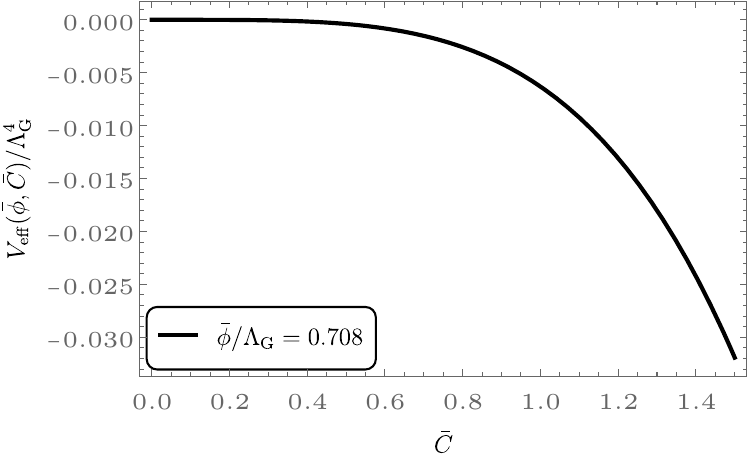}
\caption{Top: the effective potential~\eqref{eq:effective potential} at one loop as a function of $\bar\phi$ and $\bar C$. Bottom: effective potential along the $\bar\phi/\Lambda_\tx{G}$ plane (left) and the $\bar C$ plane (right).}
\label{fig:Reg effective potential}
\end{figure}

As a benchmark, we choose the following parameters:
\al{
&\xi = 0.1,&
&\bar R = 0.5\Lambda_\tx{G}^2,&
&y = 0.1,&
&\lambda = 0.6,
\label{eq:parameter choice}
}
where all dimensionful quantities are in units of the cutoff scale $\Lambda_\tx{G}$. Figure~\ref{fig:Reg effective potential} shows the effective potential for this parameter set. The bottom-left panel reveals a stable local minimum in the $\bar\phi$ direction with $\langle \bar\phi\rangle/\Lambda_\tx{G} = 0.708$. This gives the Planck mass as $M_\tx{pl} = \xi\langle \bar\phi\rangle = 0.0708\Lambda_\tx{G}$, matching the observed value with a suitable choice of $\Lambda_\tx{G}$. 

In the $\bar C$ direction, the potential diverges negatively, preventing a stable vacuum. This runaway behavior could correspond to early Universe cosmological expansion~\cite{Giddings:1991qi} or might be stabilized by subleading effects like loop fluctuations of $\phi$ and $e^\ba{}_\mu$. Crucially, the origin of the effective potential in the $\bar C$ direction is tachyonic and unstable. 

\section{Conclusion and discussion}
\label{sec:Conclusions}
We have presented a model for simultaneously generating the Planck mass scale and spacetime within the framework of classical scale-invariant degenerate gravity, incorporating the irreversible vierbein postulate. The vacuum expectation value of the scalar $\phi$ plays a central role in scalegenesis and pregeometry. The spontaneous breaking along the $\phi$ direction is triggered by the background positive curvature $\bar R$, characteristic of the LL gauge theory, which does not appear in ordinary Yang-Mills gauge theories. This nontrivial vacuum is stable against one-loop corrections from the spinor field $\psi$, the sole degree of freedom in the symmetric degenerate phase.

The conformal mode of the vierbein, $C$, acquires a nonzero vacuum expectation value through a potential of the form $\sim C^4 V(\phi)$, breaking LL gauge symmetry while maintaining general coordinate covariance. In the LL broken phase, the vierbein, LL gauge fields, and scalar $\phi$ become dynamical due to loop corrections involving $\psi$, leading to the dynamical emergence of gravity, as discussed in earlier works~\cite{Matsuzaki:2020qzf,Maitiniyazi:2023hts}.

The runaway behavior of the vierbein, illustrated in Fig.~\ref{fig:Reg effective potential} (bottom-right panel), resembles the scale factor in an expanding Universe. The scalar $\phi$ could potentially be identified with the {\it cosmon} field~\cite{Ratra:1987rm,Wetterich:1987fm}, offering avenues for exploring cosmological implications.

Our model achieves both scalegenesis and spacetime generation through symmetry breaking. The effective potential at the vacuum yields a negative cosmological constant, $\Lambda_\tx{cc} = \left.\frac{V_{\rm eff}}{C^4}\right|_{\langle \bar C\rangle \neq 0, \langle \bar \phi\rangle \neq 0} < 0$, suggesting that the generated spacetime would initially correspond to anti-de Sitter space. Resolving this discrepancy with our observed Universe would require an additional contribution to the cosmological constant. While power divergences could be considered, their dependence on regularization schemes makes this approach uncertain. As addressing the cosmological constant problem is beyond the scope of this work, we leave this issue open for future investigation.

\begin{acknowledgements}
The work of S.\,M.\ is supported in part by the National Science Foundation of China (NSFC) under Grant No.~11747308, 11975108, and 12047569, and the Seeds Funding of Jilin University. 
The work of K.\,O.\ is in part supported by JSPS KAKENHI Grant No.~19H01899.
The work of M.\,Y.\ is supported by the National Science Foundation of China (NSFC) under Grant No.~12205116 and the Seeds Funding of Jilin University.
\end{acknowledgements}
\bibliographystyle{JHEP} 
\bibliography{refs}

\providecommand{\href}[2]{#2}\begingroup\raggedright\begin{thebibliography}{10}

\bibitem{Matsuzaki:2020qzf}
S.~Matsuzaki, S.~Miyawaki, K.-y. Oda and M.~Yamada, \emph{{Dynamically emergent
  gravity from hidden local Lorentz symmetry}},
  \href{https://doi.org/10.1016/j.physletb.2020.135975}{\emph{Phys. Lett. B}
  {\bfseries 813} (2021) 135975}
  [\href{https://arxiv.org/abs/2003.07126}{{\ttfamily 2003.07126}}].

\bibitem{Maitiniyazi:2023hts}
Y.~Maitiniyazi, S.~Matsuzaki, K.-y. Oda and M.~Yamada, \emph{{Irreversible
  vierbein postulate: Emergence of spacetime from quantum phase transition}},
  \href{https://doi.org/10.1103/PhysRevD.109.106018}{\emph{Phys. Rev. D}
  {\bfseries 109} (2024) 106018}
  [\href{https://arxiv.org/abs/2309.16230}{{\ttfamily 2309.16230}}].

\bibitem{Tseytlin:1981ks}
A.~A. Tseytlin, \emph{{On the First Order Formalism in Quantum Gravity}},
  \href{https://doi.org/10.1088/0305-4470/15/3/005}{\emph{J. Phys.} {\bfseries
  A15} (1982) L105}.

\bibitem{Horowitz:1990qb}
G.~T. Horowitz, \emph{{Topology change in classical and quantum gravity}},
  \href{https://doi.org/10.1088/0264-9381/8/4/007}{\emph{Class. Quant. Grav.}
  {\bfseries 8} (1991) 587}.

\bibitem{Floreanini:1991cw}
R.~Floreanini and R.~Percacci, \emph{{Mean field quantum gravity}},
  \href{https://doi.org/10.1103/PhysRevD.46.1566}{\emph{Phys. Rev. D}
  {\bfseries 46} (1992) 1566}.

\bibitem{Wetterich:1983bi}
C.~Wetterich, \emph{{Fine Tuning Problem and the Renormalization Group}},
  \href{https://doi.org/10.1016/0370-2693(84)90923-7}{\emph{Phys. Lett. B}
  {\bfseries 140} (1984) 215}.

\bibitem{Bardeen:1995kv}
W.~A. Bardeen, \emph{{On naturalness in the standard model}},  in \emph{{Ontake
  Summer Institute on Particle Physics}}, 8, 1995.

\bibitem{Hamada:2021jls}
Y.~Hamada, H.~Kawai, K.~Kawana, K.-y. Oda and K.~Yagyu, \emph{{Minimal scenario
  of criticality for electroweak scale, neutrino masses, dark matter, and
  inflation}},
  \href{https://doi.org/10.1140/epjc/s10052-021-09735-z}{\emph{Eur. Phys. J. C}
  {\bfseries 81} (2021) 962}
  [\href{https://arxiv.org/abs/2102.04617}{{\ttfamily 2102.04617}}].

\bibitem{Kawai:2021lam}
H.~Kawai and K.~Kawana, \emph{{The multicritical point principle as the origin
  of classical conformality and its generalizations}},
  \href{https://doi.org/10.1093/ptep/ptab161}{\emph{PTEP} {\bfseries 2022}
  (2022) 013B11} [\href{https://arxiv.org/abs/2107.10720}{{\ttfamily
  2107.10720}}].

\bibitem{Hamada:2022soj}
Y.~Hamada, H.~Kawai, K.~Kawana, K.-y. Oda and K.~Yagyu, \emph{{Gravitational
  waves in models with multicritical-point principle}},
  \href{https://doi.org/10.1140/epjc/s10052-022-10440-8}{\emph{Eur. Phys. J. C}
  {\bfseries 82} (2022) 481}
  [\href{https://arxiv.org/abs/2202.04221}{{\ttfamily 2202.04221}}].

\bibitem{Kawai:2023viy}
H.~Kawai, K.~Kawana, K.-y. Oda and K.~Yagyu, \emph{{Quantum phase transition
  and absence of quadratic divergence in generalized quantum field theories}},
  \href{https://doi.org/10.1103/PhysRevD.109.085009}{\emph{Phys. Rev. D}
  {\bfseries 109} (2024) 085009}
  [\href{https://arxiv.org/abs/2307.11420}{{\ttfamily 2307.11420}}].

\bibitem{Wetterich:2016uxm}
C.~Wetterich and M.~Yamada, \emph{{Gauge hierarchy problem in asymptotically
  safe gravity--the resurgence mechanism}},
  \href{https://doi.org/10.1016/j.physletb.2017.04.049}{\emph{Phys. Lett. B}
  {\bfseries 770} (2017) 268}
  [\href{https://arxiv.org/abs/1612.03069}{{\ttfamily 1612.03069}}].

\bibitem{Kubo:2018kho}
J.~Kubo, M.~Lindner, K.~Schmitz and M.~Yamada, \emph{{Planck mass and inflation
  as consequences of dynamically broken scale invariance}},
  \href{https://doi.org/10.1103/PhysRevD.100.015037}{\emph{Phys. Rev. D}
  {\bfseries 100} (2019) 015037}
  [\href{https://arxiv.org/abs/1811.05950}{{\ttfamily 1811.05950}}].

\bibitem{Volovik:2023pcm}
G.~E. Volovik, \emph{{Planck Constants in the Symmetry Breaking Quantum
  Gravity}}, \href{https://doi.org/10.3390/sym15050991}{\emph{Symmetry}
  {\bfseries 15} (2023) 991}
  [\href{https://arxiv.org/abs/2304.04235}{{\ttfamily 2304.04235}}].

\bibitem{Edery:2015wha}
A.~Edery and Y.~Nakayama, \emph{{Generating Einstein gravity, cosmological
  constant and Higgs mass from restricted Weyl invariance}},
  \href{https://doi.org/10.1142/S0217732315501527}{\emph{Mod. Phys. Lett. A}
  {\bfseries 30} (2015) 1550152}
  [\href{https://arxiv.org/abs/1502.05932}{{\ttfamily 1502.05932}}].

\bibitem{Edery:2019txq}
A.~Edery and Y.~Nakayama, \emph{{Palatini formulation of pure $R^2$ gravity
  yields Einstein gravity with no massless scalar}},
  \href{https://doi.org/10.1103/PhysRevD.99.124018}{\emph{Phys. Rev. D}
  {\bfseries 99} (2019) 124018}
  [\href{https://arxiv.org/abs/1902.07876}{{\ttfamily 1902.07876}}].

\bibitem{Edery:2014nha}
A.~Edery and Y.~Nakayama, \emph{{Restricted Weyl invariance in four-dimensional
  curved spacetime}},
  \href{https://doi.org/10.1103/PhysRevD.90.043007}{\emph{Phys. Rev. D}
  {\bfseries 90} (2014) 043007}
  [\href{https://arxiv.org/abs/1406.0060}{{\ttfamily 1406.0060}}].

\bibitem{Ghilencea:2018dqd}
D.~M. Ghilencea, \emph{{Spontaneous breaking of Weyl quadratic gravity to
  Einstein action and Higgs potential}},
  \href{https://doi.org/10.1007/JHEP03(2019)049}{\emph{JHEP} {\bfseries 03}
  (2019) 049} [\href{https://arxiv.org/abs/1812.08613}{{\ttfamily
  1812.08613}}].

\bibitem{Ghilencea:2018thl}
D.~M. Ghilencea and H.~M. Lee, \emph{{Weyl gauge symmetry and its spontaneous
  breaking in the standard model and inflation}},
  \href{https://doi.org/10.1103/PhysRevD.99.115007}{\emph{Phys. Rev. D}
  {\bfseries 99} (2019) 115007}
  [\href{https://arxiv.org/abs/1809.09174}{{\ttfamily 1809.09174}}].

\bibitem{Shtanov:2023lci}
Y.~Shtanov, \emph{{Electroweak symmetry breaking by gravity}},
  \href{https://doi.org/10.1007/JHEP02(2024)221}{\emph{JHEP} {\bfseries 02}
  (2024) 221} [\href{https://arxiv.org/abs/2305.17582}{{\ttfamily
  2305.17582}}].

\bibitem{Karananas:2024xja}
G.~K. Karananas, M.~Shaposhnikov and S.~Zell, \emph{{Weyl-invariant
  Einstein-Cartan gravity: unifying the strong CP and hierarchy puzzles}},
  \href{https://arxiv.org/abs/2406.11956}{{\ttfamily 2406.11956}}.

\bibitem{Akama:1978pg}
K.~Akama, \emph{{An Attempt at Pregeometry: Gravity With Composite Metric}},
  \href{https://doi.org/10.1143/PTP.60.1900}{\emph{Prog. Theor. Phys.}
  {\bfseries 60} (1978) 1900}.

\bibitem{Terazawa:1980vf}
H.~Terazawa and K.~Akama, \emph{{Subquark Pregeometry With Spontaneously Broken
  Conformal Invariance}},
  \href{https://doi.org/10.1016/0370-2693(80)90551-1}{\emph{Phys. Lett. B}
  {\bfseries 97} (1980) 81}.

\bibitem{Akama:1981dk}
K.~Akama, \emph{{PREGEOMETRY INCLUDING FUNDAMENTAL GAUGE BOSONS}},
  \href{https://doi.org/10.1103/PhysRevD.24.3073}{\emph{Phys. Rev. D}
  {\bfseries 24} (1981) 3073}.

\bibitem{Akama:1979tm}
K.~Akama, \emph{{Quantum Mechanical Equivalence of Two Lagrangian Formalisms in
  'Scalar Pregeometry'}}, \href{https://doi.org/10.1143/PTP.61.687}{\emph{Prog.
  Theor. Phys.} {\bfseries 61} (1979) 687}.

\bibitem{Akama:1981kq}
K.~Akama and H.~Terazawa, \emph{{Pregeometric Origin of the Big Bang}},
  \href{https://doi.org/10.1007/BF00759207}{\emph{Gen. Rel. Grav.} {\bfseries
  15} (1983) 201}.

\bibitem{Akama:1990ur}
K.~Akama and I.~Oda, \emph{{Topological pregauge pregeometry}},
  \href{https://doi.org/10.1016/0370-2693(91)91652-C}{\emph{Phys. Lett. B}
  {\bfseries 259} (1991) 431}.

\bibitem{Akama:1991pv}
K.~Akama and I.~Oda, \emph{{Chern-Simons pregeometry}},
  \href{https://doi.org/10.1143/PTP.89.215}{\emph{Prog. Theor. Phys.}
  {\bfseries 89} (1993) 215}.

\bibitem{Akama:1991np}
K.~Akama and I.~Oda, \emph{{BRST quantization of pregeometry and topological
  pregeometry}},
  \href{https://doi.org/10.1016/0550-3213(93)90192-R}{\emph{Nucl. Phys. B}
  {\bfseries 397} (1993) 727}.

\bibitem{Akama:1994ehj}
K.~Akama and I.~Oda, \emph{{BRST quantization of pregeometry and topological
  pregeometry}}, {\emph{Adv. Appl. Clifford Algebras} {\bfseries 4} (1994)
  259}.

\bibitem{Wetterich:2021ywr}
C.~Wetterich, \emph{{Pregeometry and euclidean quantum gravity}},
  \href{https://doi.org/10.1016/j.nuclphysb.2021.115526}{\emph{Nucl. Phys. B}
  {\bfseries 971} (2021) 115526}
  [\href{https://arxiv.org/abs/2101.07849}{{\ttfamily 2101.07849}}].

\bibitem{Wetterich:2021cyp}
C.~Wetterich, \emph{{Cosmology from pregeometry}},
  \href{https://doi.org/10.1103/PhysRevD.104.104040}{\emph{Phys. Rev. D}
  {\bfseries 104} (2021) 104040}
  [\href{https://arxiv.org/abs/2104.14013}{{\ttfamily 2104.14013}}].

\bibitem{Wetterich:2021hru}
C.~Wetterich, \emph{{Pregeometry and spontaneous time-space asymmetry}},
  \href{https://doi.org/10.1007/JHEP06(2022)069}{\emph{JHEP} {\bfseries 06}
  (2022) 069} [\href{https://arxiv.org/abs/2101.11519}{{\ttfamily
  2101.11519}}].

\bibitem{Floreanini:1990cf}
R.~Floreanini and R.~Percacci, \emph{{Topological pregeometry}},
  \href{https://doi.org/10.1142/S0217732390002560}{\emph{Mod. Phys. Lett. A}
  {\bfseries 5} (1990) 2247}.

\bibitem{Volovik:2021wut}
G.~E. Volovik, \emph{{Gravity from Symmetry Breaking Phase Transition}},
  \href{https://doi.org/10.1007/s10909-022-02694-z}{\emph{J. Low Temp. Phys.}
  {\bfseries 207} (2022) 127}
  [\href{https://arxiv.org/abs/2111.07817}{{\ttfamily 2111.07817}}].

\bibitem{Ohanian:1969xhl}
H.~C. Ohanian, \emph{{Gravitons as goldstone bosons}},
  \href{https://doi.org/10.1103/PhysRev.184.1305}{\emph{Phys. Rev.} {\bfseries
  184} (1969) 1305}.

\bibitem{Nakanishi:1979fg}
N.~Nakanishi and I.~Ojima, \emph{{Proof of the Exact Masslessness of
  Gravitons}}, \href{https://doi.org/10.1103/PhysRevLett.43.91}{\emph{Phys.
  Rev. Lett.} {\bfseries 43} (1979) 91}.

\bibitem{Percacci:1990wy}
R.~Percacci, \emph{{The Higgs phenomenon in quantum gravity}},
  \href{https://doi.org/10.1016/0550-3213(91)90510-5}{\emph{Nucl. Phys. B}
  {\bfseries 353} (1991) 271}
  [\href{https://arxiv.org/abs/0712.3545}{{\ttfamily 0712.3545}}].

\bibitem{Floreanini:1991gi}
R.~Floreanini, R.~Percacci and E.~Spallucci, \emph{{Coleman-Weinberg effect in
  quantum gravity}},
  \href{https://doi.org/10.1088/0264-9381/8/9/001}{\emph{Class. Quant. Grav.}
  {\bfseries 8} (1991) L193}.

\bibitem{Floreanini:1994ypa}
R.~Floreanini and R.~Percacci, \emph{{A Multiplicative background field
  method}}, .

\bibitem{Floreanini:1995ie}
R.~Floreanini and R.~Percacci, \emph{{Quantum mechanical breaking of local
  GL(4) invariance}},
  \href{https://doi.org/10.1016/0370-2693(96)00424-8}{\emph{Phys. Lett. B}
  {\bfseries 379} (1996) 87}
  [\href{https://arxiv.org/abs/hep-th/9508157}{{\ttfamily hep-th/9508157}}].

\bibitem{Percacci:2009ij}
R.~Percacci, \emph{{Gravity from a Particle Physicists' perspective}},
  \href{https://doi.org/10.22323/1.081.0011}{\emph{PoS} {\bfseries ISFTG}
  (2009) 011} [\href{https://arxiv.org/abs/0910.5167}{{\ttfamily 0910.5167}}].

\bibitem{Tomboulis:2011qh}
E.~T. Tomboulis, \emph{{General Relativity as the effective theory of GL(4,R)
  spontaneous symmetry breaking}},
  \href{https://doi.org/10.1103/PhysRevD.84.084018}{\emph{Phys. Rev. D}
  {\bfseries 84} (2011) 084018}
  [\href{https://arxiv.org/abs/1105.5848}{{\ttfamily 1105.5848}}].

\bibitem{Hartle:1983ai}
J.~B. Hartle and S.~W. Hawking, \emph{{Wave Function of the Universe}},
  \href{https://doi.org/10.1103/PhysRevD.28.2960}{\emph{Phys. Rev. D}
  {\bfseries 28} (1983) 2960}.

\bibitem{Halliwell:1988ik}
J.~J. Halliwell and J.~Louko, \emph{{Steepest Descent Contours in the Path
  Integral Approach to Quantum Cosmology. 1. The De Sitter Minisuperspace
  Model}}, \href{https://doi.org/10.1103/PhysRevD.39.2206}{\emph{Phys. Rev. D}
  {\bfseries 39} (1989) 2206}.

\bibitem{Vilenkin:1994rn}
A.~Vilenkin, \emph{{Approaches to quantum cosmology}},
  \href{https://doi.org/10.1103/PhysRevD.50.2581}{\emph{Phys. Rev. D}
  {\bfseries 50} (1994) 2581}
  [\href{https://arxiv.org/abs/gr-qc/9403010}{{\ttfamily gr-qc/9403010}}].

\bibitem{Coleman:1973jx}
S.~R. Coleman and E.~J. Weinberg, \emph{{Radiative Corrections as the Origin of
  Spontaneous Symmetry Breaking}},
  \href{https://doi.org/10.1103/PhysRevD.7.1888}{\emph{Phys. Rev. D} {\bfseries
  7} (1973) 1888}.

\bibitem{Floreanini:1993na}
R.~Floreanini and R.~Percacci, \emph{{Average effective potential for the
  conformal factor}},
  \href{https://doi.org/10.1016/0550-3213(95)00479-C}{\emph{Nucl. Phys. B}
  {\bfseries 436} (1995) 141}
  [\href{https://arxiv.org/abs/hep-th/9305172}{{\ttfamily hep-th/9305172}}].

\bibitem{Oda:2024nda}
I.~Oda, \emph{{The effective potential for conformal factor and GL(4)
  symmetry}}, \href{https://doi.org/10.1088/1572-9494/ad7832}{\emph{Commun.
  Theor. Phys.} {\bfseries 77} (2025) 015202}
  [\href{https://arxiv.org/abs/2401.04712}{{\ttfamily 2401.04712}}].

\bibitem{Giddings:1991qi}
S.~B. Giddings, \emph{{Spontaneous breakdown of diffeomorphism invariance}},
  \href{https://doi.org/10.1016/0370-2693(91)90915-D}{\emph{Phys. Lett. B}
  {\bfseries 268} (1991) 17}.

\bibitem{Wetterich:2019zdo}
C.~Wetterich and M.~Yamada, \emph{{Variable Planck mass from the gauge
  invariant flow equation}},
  \href{https://doi.org/10.1103/PhysRevD.100.066017}{\emph{Phys. Rev. D}
  {\bfseries 100} (2019) 066017}
  [\href{https://arxiv.org/abs/1906.01721}{{\ttfamily 1906.01721}}].

\bibitem{Vassilevich:2003xt}
D.~V. Vassilevich, \emph{{Heat kernel expansion: User's manual}},
  \href{https://doi.org/10.1016/j.physrep.2003.09.002}{\emph{Phys. Rept.}
  {\bfseries 388} (2003) 279}
  [\href{https://arxiv.org/abs/hep-th/0306138}{{\ttfamily hep-th/0306138}}].

\bibitem{Ratra:1987rm}
B.~Ratra and P.~J.~E. Peebles, \emph{{Cosmological Consequences of a Rolling
  Homogeneous Scalar Field}},
  \href{https://doi.org/10.1103/PhysRevD.37.3406}{\emph{Phys. Rev. D}
  {\bfseries 37} (1988) 3406}.

\bibitem{Wetterich:1987fm}
C.~Wetterich, \emph{{Cosmology and the Fate of Dilatation Symmetry}},
  \href{https://doi.org/10.1016/0550-3213(88)90193-9}{\emph{Nucl. Phys. B}
  {\bfseries 302} (1988) 668}
  [\href{https://arxiv.org/abs/1711.03844}{{\ttfamily 1711.03844}}].

\end{thebibliography}\endgroup
\end{document}